\documentstyle[12pt,epsf]{article} 

\makeatletter

\def\vereq#1#2{\lower3pt\vbox{\baselineskip1.5pt \lineskip1.5pt
\ialign{$\m@th#1\hfill##\hfil$\crcr#2\crcr\sim\crcr}}}
\def\gtrsim{\mathrel{\mathpalette\vereq>}}

\def\agt{\gtrsim}
\makeatother

\title{
{\normalsize E-print hep-ph/9812391 \hfill Preprint YARU-HE-98/07} \\[7mm]
{\LARGE\bf Field-induced axion emission via process 
                 $e^+ e^- \to a$ in plasma} 
}
\author{\large N.V.~Mikheev, A.Ya.~Parkhomenko and L.A.~Vassilevskaya \\ 
        {\normalsize\it Yaroslavl State (Demidov) University,} \\ 
        {\normalsize\it Sovietskaya 14, Yaroslavl 150000, Russia}} 

\date{} 

\begin{document} 

\maketitle 

\begin{abstract} 
The annihilation into axion $e^+ e^- \to a$ is investigated in a plasma 
and an external magnetic field. This process via a plasmon intermediate 
state has a resonant character at a particular energy of the emitted 
axion. The emissivity by $e^+ e^- \to a$ is compared with the axion 
cyclotron emissivity. 
\end{abstract} 


\section{Introduction} 

The axion~\cite{Peccei77,WW}, the pseudo-Goldstone boson arising as 
a result of the spontaneous breakdown of the global 
Peccei-Quinn (PQ) symmetry $U_{PQ} (1)$, is one of the 
well-motivated candidates for the cold dark matter (see, for example, 
Refs.~\cite{Raffelt-book,Raffelt-castle97,Raffelt-school97} 
and references therein). 
Although the original axion, associated with the PQ symmetry 
breakdown at the weak scale ($f_w$), is excluded experimentally, many 
variants of the PQ models and their accompanying axions are of great 
interest. If the breaking scale of the PQ symmetry $f_a$ is much larger 
than the electroweak scale $f_a \gg f_w$ (the latest 
astrophysical data yield $f_a \agt 10^{10}$~GeV), the resulting 
``invisible axion''~\cite{KSVZ,DFSZ} is very light ($m_a \sim f_a^{-1}$) 
and very weakly coupled (coupling $\sim f_a^{-1}$). 

In view of the smallness of the coupling constant, axion effects could 
be noticeable under astrophysical conditions -- high matter densities, 
high temperature, and strong magnetic fields. So, it is important 
to take into account the influence of a plasma and magnetic fields in 
studies of axion processes in stars. One of the most physically realistic 
situations presented in many astrophysical objects is that when from 
both these components of the active medium the plasma dominates:  
\begin{equation} e B \ll \mu^2, T^2, 
\label{eq:condition}  
\end{equation}  
where $\mu$ and $T$ are the electron chemical potential and temperature, 
respectively. While this situation corresponds to a relatively weak 
magnetic field, it can still be strong $B \gg B_e$ in comparison with 
the electron's Schwinger value 
$B_e = m^2_e/e \simeq 4.41 \times 10^{13}$~G. 
At present an existence of magnetic fields up to 
$B \sim 10^{15} - 10^{17}$~G 
is not exotic in astrophysics~\cite{magnetar,toroidal} and 
cosmology~\cite{cosmol}. A possible manifestation of 
axions' effects under such conditions is of great interest. 

In the recent paper~\cite{MRV} the axion cyclotron emission of the 
plasma $e^- \to e^- a$ was studied as a possible source of energy losses 
by astrophysical objects.  In this paper we study the annihilation process 
$e^+ e^- \to a$ in medium as an additional channel of energy losses and 
compare the emissivity by this process with the axion cyclotron emissivity  
under the condition of a nondegenerate hot plasma. 

\section{$S$-matrix element} 

\begin{figure}[tb]  
\centerline{\epsfxsize=.80\textwidth \epsffile[95 540 520 690]{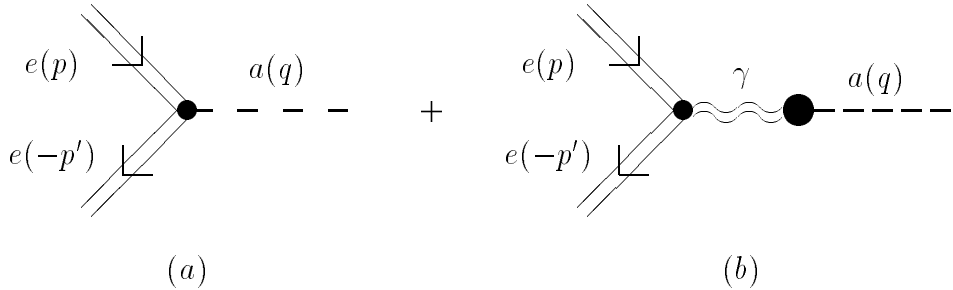}} 
\caption{The annihilation process.} 
\label{fig:eea} 
\end{figure}

The process $e^+ e^- \to a$ is described by two diagrams in 
Fig.~\ref{fig:eea} 
%
%
%
%
where solid double lines imply the influence of medium in the electron 
and positron wave functions and the photon propagator. 
Figure~\ref{fig:eea}a  describes the axion annihilation 
of $e^+ e^-$-pair due to a direct axion-fermion coupling: 
\begin{equation} 
{\cal L}_{af} = - i g_{af}\;(\bar f \gamma_5 f) \, a, 
\label{eq:L-aff} 
\end{equation} 
where $g_{af} = C_f m_f/f_a$ is a dimensionless Yukawa coupling constant; 
$C_f$ is a model-dependent factor~\cite{DFSZ}, 
$m_f$ is the fermion mass (the electron mass in our case); 
$f$ and $a$ are the fermion and axion fields, respectively.  

\begin{figure}[tb] 
\centerline{\epsfxsize=.80\textwidth 
            \epsffile[135 520 455 715]{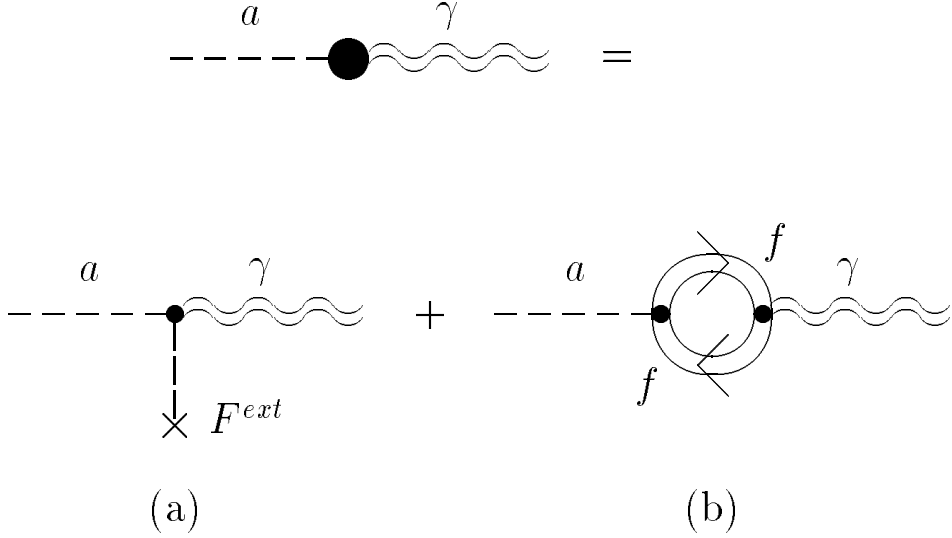}} 
\caption{Axion-photon coupling in an external electromagnetic field.} 
\label{fig:ag-eff} 
\end{figure}

Figure~\ref{fig:eea}b 
describes the electron-positron annihilation via the photon intermediate 
state. This channel becomes possible due to an effective axion-photon 
interaction with the Lagrangian: 
\begin{eqnarray} 
{\cal L}_{a \gamma} = \bar g_{a\gamma} \, (\partial_\mu A_\nu) \, 
\tilde F^{\nu\mu} \, a, 
\label{eq:Lag} 
\end{eqnarray} 
where $A_\mu$ is the four-potential of the quantized electromagnetic 
field; $\tilde F_{\mu\nu}$ is the dual tensor of the external field;  
$\bar g_{a\gamma}$ is an effective coupling in the presence 
of the magnetic field with the dimension $(energy)^{-1}$~\cite{MRV}:  
\begin{eqnarray} 
\bar g_{a\gamma} = g_{a\gamma} + \Delta g_{a \gamma}. 
\label{eq:Gag1} 
\end{eqnarray} 
Here $g_{a \gamma}$ corresponds to the well-known 
$a\gamma\gamma$ coupling in vacuum (Fig.~\ref{fig:ag-eff}a) 
%
%
%
with the constant $g_{a \gamma} = \alpha \xi/ 2 \pi f_a $~\cite{Raffelt90},  
where $\xi$ is the model-dependent parameter. 
The second term in Eq.~(\ref{eq:Gag1}) 
is the field-induced contribution to the effective axion-photon coupling 
$\bar g_{a\gamma}$ which comes from Fig.~\ref{fig:ag-eff}b: 
\begin{eqnarray} 
\Delta g_{a\gamma} & = & \frac{\alpha}{\pi} \; 
\sum_f \frac{Q_f^2 g_{af}}{m_f} \; (1 - J) , 
\label{eq:J} \\ 
J & = & \left ( {4 \over \chi_f} \right )^{2/3} \, 
\int \limits_0^{\pi/2} \; f(\eta) \, \sin^{- 1/3} \phi \, d\phi,
\nonumber 
\end{eqnarray} 
where the Hardy-Stokes function $f (\eta)$ of the argument 
$\eta = (4/\chi_f \sin^2 \phi)^{2/3}$ is defined as:  
\begin{eqnarray} 
f(\eta) = i\,\int \limits_0^\infty du \, \exp \left \lbrace -i \, 
\left ( \eta u + { u^3 \over 3 } \right ) \right \rbrace.  
\nonumber 
\end{eqnarray} 
Further, the dynamic parameter is $\chi_f^2 = e_f^2 (qFFq)/m_f^6$,   
where $e_f = e Q_f$, $e>0$ is the elementary charge, 
$Q_f$ is a relative electric charge of a loop fermion,  
$(qFFq) = q_\mu F^{\mu\nu} F_{\nu\rho} q^\rho$, 
$q_\mu = (E_a,{\bf q})$ is the four-momentum of the final axion, 
and $F_{\mu \nu}$ is the external electromagnetic field tensor. 
The field-induced coupling $\Delta g_{a\gamma}$ 
is a step function of the magnetic field 
strength because of the hierarchy of the fermion mass spectrum.  

The case of ultrarelativistic electrons and the relatively weak external 
magnetic field, when a large number of the Landau levels is excited, 
is well described by a crossed field limit (${\bf E} \perp {\bf B}$, 
$E = B$), and the dynamic parameter $\chi_f$ is the only field invariant. 

The matrix element of the process $e^-(p) + e^+(p') \to a(q)$ 
shown in Fig.~\ref{fig:eea} is the sum: 
\begin{equation} 
S = S^{(a)} + S^{(b)}, 
\label{eq:S1} 
\end{equation} 
where $S^{(a)}$ corresponds to Fig.~\ref{fig:eea}a: 
\begin{eqnarray} 
S^{(a)} = \frac{g_{ae}}{\sqrt{2 E_a V}} \;\int d^4 x\, 
{\bar\psi(-p',x)}\,\gamma_5\,\psi(p,x) \, e^{iqx}, 
\nonumber  
\end{eqnarray} 
and $S^{(b)}$ describes the contribution from Fig.~\ref{fig:eea}b: 
\begin{eqnarray} 
S^{(b)} & = & \frac{\bar g_{a\gamma}}{\sqrt{2 E_a V}} \int d^4 x \, 
{\bar\psi(-p',x)} \, (\gamma h) \, \psi(p,x) \, e^{iqx}, 
\nonumber \\ 
h_\alpha & = & 
- i e (q \tilde F G^L (q))_\alpha = 
- i e q_\mu \tilde F^{\mu\nu} G^L_{\nu\alpha} (q). 
\nonumber 
\end{eqnarray} 
Here, $\psi(p,x)$ is the exact solution of the Dirac equation in the  
external crossed field~\cite{BLP}; $p_\mu = (E, {\bf p})$ and 
${p'}_\mu =(E', {\bf p}')$ are the four-mo\-men\-ta of the initial 
electron and positron ($p^2 = {p'}^2 = m_e^2$); 
$(\gamma h) = \gamma_\mu h^\mu$, 
$\gamma_\mu$ are the Dirac $\gamma$-matrices; 
$V$ is the three-dimensional volume. 
 
Note that from both components of an active medium, the plasma determines 
basically the properties of the photon (plasmon) propagator. The 
contribution to the amplitude from the transverse intermediate plasmons 
is negligible small in the ultrarelativistic limit. 
Finally, $G^{L}_{\alpha\beta}$ is the longitudinal plasmon propagator: 
\begin{equation} 
G^{L}_{\alpha\beta} = i \, \frac{\ell_\alpha \ell_\beta}{q^2 - \Pi^L}. 
\label{eq:G} 
\end{equation} 
Here, $\Pi^{L}$ is the eigenvalue of the polarization operator of the  
longitudinal plasmon with the eigenvector: 
\begin{equation} 
\ell_\alpha = \sqrt{\frac{q^2}{(uq)^2 - q^2}} \, 
\left ( u_\alpha - \frac{uq}{q^2} \, q_\alpha \right ) , 
\nonumber 
\end{equation} 
where $u_{\alpha}$ is the four-velocity of medium. 

By integrating over the variable $x$, Eq.~(\ref{eq:S1}) can be 
presented in the ultrarelativistic limit of the form: 
\begin{eqnarray} 
S & = & {(2 \pi)^4 \delta^{(2)} ({\bf Q}_{\perp}) \; \delta(k Q) 
\over \sqrt {2 E_a V \cdot 2 E V \cdot 2 E' V}} \; 
\frac{\Phi(\eta)}{\pi \ae z}\; 
\label{eq:S2} \\ 
& \times & 
\bar U(-p') \, \bigg [ \;  
g_{ae} \gamma_5 \left ( 1 - \frac{i e z^2}{m^2_e} \; 
(\gamma k) (\gamma a) \; \frac{\Phi'(\eta)}{\Phi(\eta)} \right ) 
+ \bar g_{a \gamma} \; (\gamma h) \; \bigg ] \; U (p), 
\nonumber \\ 
\ae^2 & = & - \frac{e^2 a^2}{m_e^2}, \qquad  
z = \left (\frac{\chi_a}{2 \chi \chi'} \right )^{1/3}, 
\nonumber \\ 
\chi^2 & = & \frac{e^2 (p F F p)}{m_e^2} , \qquad 
\chi_a^2  = \frac{e^2 (q F F q)}{m_e^2} , \qquad 
\chi' = \chi (p \to p') . 
\nonumber 
\end{eqnarray} 
Here, $F_{\mu\nu} = k_\mu a_\nu - k_\nu a_\mu$ ($k^2 = (ka) = 0$) 
is the external crossed field tensor; $Q = q - p - p'$, 
${\bf Q}_\perp$ is the perpendicular to $\bf k$ component 
(${\bf Q}_\perp {\bf k} = 0$). 
The bispinor $U (p)$, which is normalized by the condition 
$\bar U U = 2 m_e$, satisfies the Dirac equation for the free electron 
$((\gamma p) - m_e) U(p) = 0$. 
Finally, $\Phi (\eta)$ is the Airy function: 
\begin{eqnarray}
\Phi (\eta) & = &  \int\limits_0^\infty d t \cos \left ( \eta t +
{t^3\over 3} \right ),  
\nonumber \\ 
\eta & = & z^2 \; (1 + \tau^2) , \qquad 
\tau = - \, \frac{e (p \tilde F q)}{m_e^4 \chi_a} ,
\nonumber 
\end{eqnarray}
and $\Phi' (\eta) = d \Phi (\eta)/ d \eta$. 

\section{Axion emissivity} 

\begin{figure}[tb] 
\centerline{\epsfxsize=.60\textwidth 
            \epsffile[115 430 395 710]{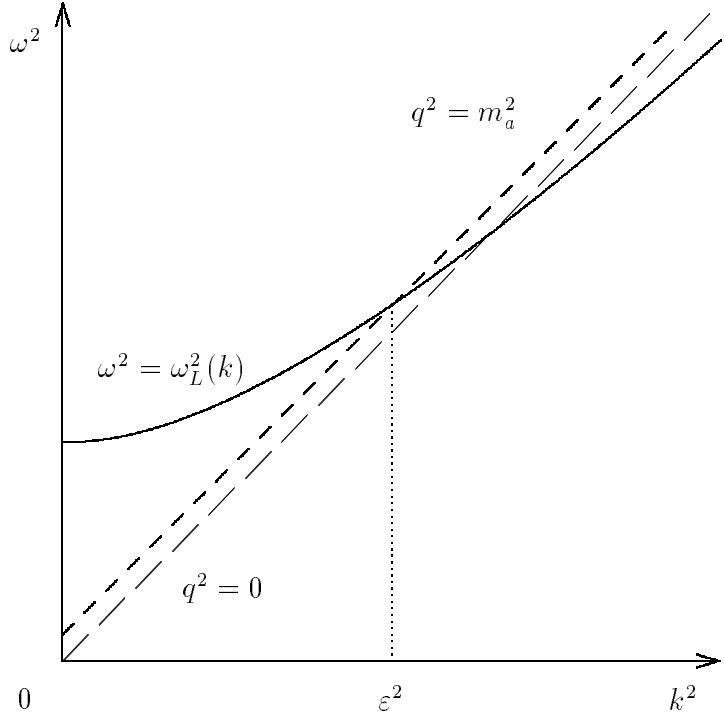}} 
\caption{Dispersion relations $\omega^2 = \omega^2_L(k)$ for 
         longitudinal plasmons (solid line), axions  
         $E_a^2 = k^2 + m^2_a$ (short dashes), and vacuum photons 
         $\omega = k$ (long dashes).} 
\label{fig:disp} 
\end{figure}

The plasma's axion emissivity due 
to the process $e^+ e^- \to a $ can be written as: 
\begin{equation} 
Q_a = \frac{1}{V T_0} \; \int d n_{e^-} \int d n_{e^+} 
\int \frac{V\; d^3 {\bf q}}{(2 \pi)^3} \;E_a \; \sum_{s,s'} | S |^2, 
\label{eq:Q1} 
\end{equation} 
where $T_0$ is a time interval. 
Taking into account the condition~(\ref{eq:condition}), the numbers 
of plasma electrons and positrons can be estimated as the numbers 
without the field: 
\begin{equation} 
d n_{e^-} \simeq \frac{V \; d^3 {\bf p}}{(2 \pi)^3} \; f (E) , \quad 
d n_{e^+} \simeq \frac{V \; d^3 {\bf p'}}{(2 \pi)^3} \; \bar f (E') ,  
\label{eq:ep-numbers} 
\end{equation}  
where $f(E)=(e^{(E - \mu)/T} + 1)^{-1}$ and 
$\bar f(E')=(e^{(E' + \mu)/T} + 1)^{-1}$ are the electron's and 
positron's Fermi-Dirac distribution functions at the temperature~$T$ 
and the chemical potential~$\mu$, respectively. 
Carrying out the integration in Eq.~(\ref{eq:Q1}), we obtain the  
expression for the axion emissivity in the form: 
\begin{eqnarray} 
\frac{d Q_a}{d E_a} & \simeq &
\frac{g^2_{ae} \; (e B)^{2/3}}{40 \pi^{5/2} 3^{1/3} \Gamma(5/6)} \; 
E_a^{7/3} \; I_{- \frac{1}{3}, - \frac{1}{3}} (E_a) 
\label{eq:Q2} \\ 
& + & 
\frac{\bar g^2_{a\gamma} \; (e B)^2}{36 \pi^3} \; 
\frac{E_a^3}{({\cal E}^2 - E_a^2)^2 + \gamma^2 {\cal E}^4} \, 
I_{1,1} (E_a) , 
\nonumber  
\end{eqnarray} 
where 
\begin{eqnarray} 
I_{k,n} (E_a) = \int^{\infty}_{E_a} dE \, E^k \, (E_a - E)^n \, 
f (E) \, \bar f (E_a - E). 
\nonumber 
\end{eqnarray} 
The second term in Eq.~(\ref{eq:Q2}) describing the contribution of 
the longitudinal plasmon intermediate state has a resonant character  
at a particular energy of the emitted axion $E_a \sim {\cal E}$. 
This is due to the fact that the axion and the longitudinal plasmon 
dispersion relations cross for a certain wave number $k = {\cal E}$ 
as was shown in Fig.~\ref{fig:disp}. 
%
%
%
%
The dimensionless resonance width~$\gamma$ in Eq.~(\ref{eq:Q2}) is 
\begin{equation} 
\gamma = \frac{{\cal E} \Gamma_L({\cal E})}{q^2} \; 
\left ( 1 - \frac{\partial \,\Pi^{(L)}}{\partial\,q_0^2 } \right ). 
\label{eq:Gamma1} 
\end{equation} 
Here, $\Gamma_L ({\cal E})$ is the total width of the longitudinal 
plasmon in the presence of the magnetic field,  
determined mainly by the absorption $\gamma_L e^- \to e^-$~\cite{MRV}. 
The expression in brackets in Eq.~(\ref{eq:Gamma1}) 
comes from the renormalization of the longitudinal plasmon wave function. 
%
%
\begin{figure}[tb] 
\centerline{\epsfxsize=.60\textwidth 
            \epsffile[140 360 490 735]{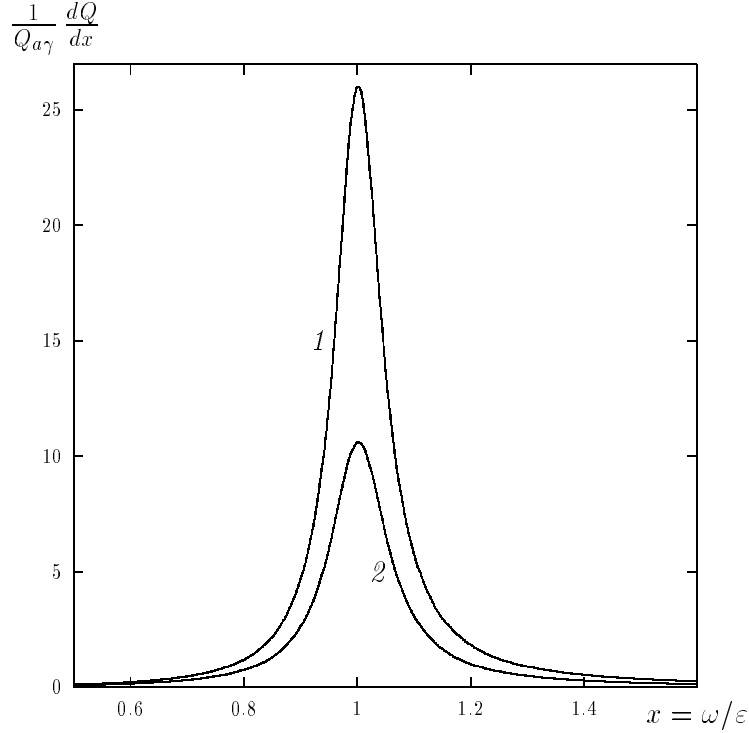}} 
\caption{ 
The resonant behaviour of the axion emissivity spectrum where 
$Q_{a\gamma} = (\bar g^2_{a\gamma}/36 \pi^3)(e B)^2\mu^2 {\cal E}$. 
Line~1 corresponds to 
$\mu = 100 \, m_e$, $T = 10 \, m_e$, ${\cal E} = 20 \, m_e$,  
while line~2 corresponds to 
$\mu = 500 \, m_e$, $T = 70 \, m_e$, ${\cal E} = 120 \, m_e$. 
} 
\label{fig:reson} 
\end{figure} 
%
%
The second term in Eq.~(\ref{eq:Q2}) defines the main contribution to 
the plasma's axion emissivity by the annihilation $e^+ e^- \to a$ in 
the resonant region. So, below we neglect the nonresonant direct 
axion-electron term. The resonant behaviour of the axion spectrum 
is shown in 
Fig.~\ref{fig:reson}. 
The expressions for ${\cal E}$, 
$\gamma$ and the axion emissivity in the resonant point are: 
\newline 
i) degenerate plasma ($\mu \gg T$) 
\begin{eqnarray} 
{\cal E}^2 & \simeq & \frac{4 \alpha}{\pi} \, \mu^2 \, 
\left ( \ln \frac{2 \mu}{m_e} - 1 \right ) , \qquad 
\gamma \simeq \frac{2 \alpha}{3} \, \frac{\mu^2}{{\cal E}^2}, 
\label{eq:degen} \\ 
Q_a^{\rm res} & \simeq &  
\frac{\bar g^2_{a\gamma} \, (e B)^2}{48 \pi^2 \alpha} \; 
\frac{{\cal E}^3 T^2}{\mu^2} \; e^{- \mu / T} , 
\nonumber  
\end{eqnarray} 
ii) nondegenerate hot plasma ($T \gg \mu$) 
\begin{eqnarray} 
{\cal E}^2 & \simeq & \frac{4 \pi \alpha}{3} \, T^2 \, 
\left ( \ln \frac{4 T}{m_e} - 0.647 \right ), \qquad 
\gamma \simeq \frac{2 \pi^2 \alpha}{9} \, \frac{T^2}{{\cal E}^2}, 
\label{eq:nondegen} \\  
Q_a^{\rm res} & \simeq &  
\frac{\bar g^2_{a\gamma} \, (e B)^2}{384 \pi^4 \alpha} \; 
\frac{{\cal E}^5}{T^2}. 
\nonumber  
\end{eqnarray} 

The conditions of the nondegenerate hot plasma are the most favourable 
for a possible manifestation of the annihilation $e^+ e^- \to a$. 
Below we compare the axion emissivity 
by the process $e^+ e^-\to a$ with the one by the axion cyclotron 
emission~\cite{MRV} under these conditions: 
\begin{eqnarray} 
\frac{Q (e^+ e^- \to a)}{Q (e^- \to e^- a)} \simeq  
\frac{1}{8 \pi^2} \; \frac{{\cal E}^3}{T^3} \sim  
\left [ \frac{\alpha}{3 \pi^{1/3}} 
\left ( \ln \frac{4T}{m_e} - 0.647 \right) \right ]^{3/2} . 
\label{eq:Rel} 
\end{eqnarray} 
It is seen from Eq.~(\ref{eq:Rel}) that the axion emissivity by the  
annihilation is noticeably smaller than that by the cyclotron emission. 
	
\section*{Acknowledgements} 

This work was partially supported by INTAS under grant No.~96-0659  
and by the Russian Foundation for Basic Research  
under grant No.~98-02-16694. 
The work of N.V.M. was supported under grant No.~d98-181 
by International Soros Science Education Program. 


\end{document}